\documentclass[11pt]{article}
\usepackage{latexsym}  
\usepackage{amssymb}
\usepackage{graphicx}
\usepackage{amsmath}

\begin{document}
\begin{center}
{\Large\bf  
  Composite Model of Quarks and Leptons 
}\\[0.1in]

Yoshio Koide

 {\it Department of Physics, Osaka University, 
Toyonaka, Osaka 560-0043, Japan} \\
{\it E-mail address: koide@epp.phys.sci.osaka-u.ac.jp}

\end{center}

\begin{quotation} 
A compsite model of quarks and leptons is proposed.
The quarks and leptons are given by three body states
which are composed of constituents
$(w_1, w_2, c_1, c_2, c_3)$ of SU(5)$_{flavor}$ and
$(f_1, f_2, f_3)$ of SU(3)$_{family}$
\end{quotation}

\vspace{5mm}

{\large\bf 1. \ Introduction } 

\vspace{2mm}

Nowadays, we know that hadrons (baryons and mesons) are
composite states composed of quarks\cite{quark}, 
so that "elementary particles" are quarks and leptons. 

 We know that quarks and leptons are described as 
$(\overline{{\bf 10}} + {\bf 5})$ of $SU(5)_{GUT}$
\cite{SU5}:
$$
{\bf 5}_L = \left( 
\begin{array}{l}
\nu_L \\
e_L \\
\overline{d_R}
\end{array} 
\right), \ \ \ \ \ \ 
\overline{\bf 10}_L = \left( 
\begin{array}{l}
\overline{e_R} \\
u_L \\
d_L \\
\overline{u_R}
\end{array}
\right). 
\eqno(1.1)
$$

If we introduce a family symmetry SU(3), 
as we discuss in Sec.2, it is 
suggested  that quarks and leptons are described by
a symmetry SU(5)$_{flavor} \times$ SU(3)$_{family}$ 
as $({\bf 5}, \overline{\bf 3})$ and 
$({\bf 10}, {\bf 3})$. 

A model with such a symmetry 
SU(5)$_{flavor} \times$ SU(3)$_{family}$ 
has been discussed by the author already \cite{YK}. 
However, since the model has been discussed 
only on the base of the symmetry, some unwelcome states 
appeared in addition to the desirable quarks and leptons.

In this paper, we discuss a possibility that 
the quarks and leptons are not elementary particles, 
but they are composed of more fundamental constituents 
("preons")\cite{preon}. 
By introducing some dynamical constraint for the  
composite states, we will remove unwelcome states 
from all the combinations of preons except for 
the desirable quark and lepton states. 

\vspace{5mm}

{\large\bf 2. \ Preons and composite states} \ 

\vspace{2mm}

We know that quarks and leptons are described as 
$(\overline{{\bf 10}} + {\bf 5})$ of $SU(5)_{GUT}$: 
$$
{\bf 5} = \left( (\nu_L, e_L), \overline{e_R} \right) , 
\ \ \ \ \ 
{\bf 10} = \left( e_R, \overline{(u_L, d_L)} , u_R \right) .
\eqno(2.1)
$$
If we denote the SU(5) symmetric states by the Young 
diagrams, the state $\bf{ 10 }$  
can be given by a two box state
which is arranged  longitudinally 
$$
{\bf 10 } =  
\begin{tabular}{c}
\fbox{ {\bf 5} } \\
\fbox{ {\bf 5} } 
\end{tabular} .
\eqno(2.2)
$$

We define more fundamental constituents (preons)
$$
{\bf 5} = (w_1, w_2, c_1, c_2, c_3) ,
\eqno(2.3)
$$
where $(w_1, w_2)$ is a doublet of SU(2)$_L$ and 
$(c_1, c_2, c_3)$ is a triplet of SU(3)$_{coloar}$. 
Then, we can describe quarks and leptons as 
$$
\fbox{ {\bf 5} }_{\ L}+ \left(
\begin{tabular}{c}
\fbox{ {\bf 5} } \\
\fbox{ {\bf 5} } 
\end{tabular} \right)_R
 ,
\eqno(2.4)
$$
i.e.
the quarks and leptons can be described by
the composite states (2.4) of preons (2.3).
Here, the assigment of the chirality ($L$ and $R$)
was chosen from a phenomenological point of view.
The rule of the chirality will be discussed in the 
next section.

The relation (2.4) suggests that if we introduce 
additional preons
$$
{\bf 3} = \left(f_1, f_2, f_3 \right) , 
\eqno(2.5)
$$
of SU(3) family, quarks and leptons 
can be regarded as three-body composite states.  
That is, we can regard quarks and leptons 
with three families as the three body 
composite states of the preons 
$$
 \left(
\begin{tabular}{c}
\fbox{ {\bf 5} } \\
\fbox{ {\bf 5} } \\
\fbox{ {\bf 3} }
\end{tabular} \right)_R
+ 
 \left(
\begin{tabular}{c}
\fbox{ {\bf 5} } \\
\fbox{ {\bf 3} } \\
\fbox{ {\bf 3} }
\end{tabular} \right)_L 
= ({\bf 10}, {\bf 3})_R +(
{\bf 5}, \overline{\bf 3} )_L .
\eqno(2.6)
$$
Here, we assume that the three ppreon states 
are bounded by the SU(3) hyper-color force. 

Of course. we must discuss further three-body 
bound states $\{ \bf{5}, \bf{5}, \bf{5} \}$ 
and $ \{ \bf{3}, \bf{3}, \bf{3}\}$.
This problem will be discussed in the next  
section Sec.3. 
The problem on the choice $L$ or $R$ will be 
also discussed in the next section. 

Lastly, let us summarize  the quantum numbers of 
preons in Table 1 and their composite states 
in Table 2.  
In both Tables 1 and 2, we show the electric 
charge $Q= I_3 +Y/2$ insteaed of the hypercharge 
$Y$. 

\begin{table}
\begin{center}
\begin{tabular}{|c|cc|} \hline
{\rm preon} &  $I_3$ & $Q$ \\ \hline 
 $w_1$ & $ +\frac{1}{2}$ & 0 \\
 $w_2$ & $-\frac{1}{2}$ & -1 \\
 $c_a$ & 0 & $+\frac{1}{3}$ \\
 $f_i$ & 0 & 0 \\ \hline
 \end{tabular}
\end{center}

 \caption{{\bf  Quantum numbers of preons}
 Here,  $a=1, 2, 3$ (color numbers) and 
 $i=1, 2, 3$ (family numbers).} 
 \end{table}
 
 \begin{table}
 \begin{center}
\begin{tabular}{|c| c| c|c|} \hline
composite state &  $I_3$ & $Q$ & $q$ or $l$ \\ \hline
 $(w_1 f_j f_k )$ &  $+ \frac{1}{2}$ & 0 & $(\nu_L)^i$ \\
 $(w_2 f_j f_k )$ & $- \frac{1}{2}$ & $-1$ & $(e_L)^i$ \\
 $(c_a f_j f_k )$ &  0 & $+\frac{1}{3}$ & 
$(\overline{d_R})_a^i$ \\ 
 $(w_1 w_2 f_i )$ &  0 & $-1$ & $(e_R)_i$ \\
 $(w_1 c_a f_i )$ &  $+ \frac{1}{2}$ & 
 $+ \frac{1}{3}$ & $\overline{(d_L)}_{ai}$ \\
 $(w_2 c_a f_i)$  &  $- \frac{1}{2}$ & 
 $- \frac{2}{3}$ & $\overline{(u_L)}_{ai} $ \\
$(c_b c_c f_i)$ & 0 & $+ \frac{2}{3}$ & 
$(u_R)^a_i $ \\ \hline
 \end{tabular} 
 
 \caption{ \bf  Quantum numbers of composite states} 
 \end{center} 
 \end{table}

\newpage

{\large\bf 3. Chirality selection rule}

\vspace{2mm}

Now, we must discuss "chirality" in the composite 
states.   However, rigorously speaking, there is 
no theory on the chirarity for composite states. 

In this paper, we put an  empirical rule of the 
chirality on the composite states as follows:

\noindent
(i) Chirality is defined for the two states $L$ and $R$,
so that the state $L$ or $R$ will be given by two
quantum numbers $\chi = -\frac{1}{2}$ and 
$\chi = + \frac{1}{2}$ analogously to two components in  
spin $J=\frac{1}{2}$ particle.
We define the chiral qauntum number as follows:
$$
\chi({\bf 5}_R) = + \frac{1}{2}, \ \ \ \ 
\chi({\bf 3}_L) = - \frac{1}{2}. 
\eqno(3.1)
$$

\noindent
(ii) The chirality of a composite state is given by
$$
\chi_{comp} = \sum \chi_{preon} .   \eqno(3.2)
$$

Then, the chirality of composite state is given by
$$
\chi({\bf 5}_R {\bf 5}_R {\bf 3}_L ) = + \frac{1}{2} , 
\ \ \ \  
\chi({\bf 5}_R {\bf 3}_L {\bf 3}_L ) = - \frac{1}{2} , 
\eqno(3.3)
$$
so that we obtain
$$
({\bf 5}_R {\bf 5}_R {\bf 3}_L )_R, \ \ \ \ 
({\bf 5}_R {\bf 3}_L {\bf 3}_L )_L . 
\eqno(3.4)
$$

On the other hand, for the composite states
$$
({\bf 5}_R {\bf 5}_R {\bf 5}_R ), \ \ \  
({\bf 3}_L {\bf 3}_L {\bf 3}_L ), 
\eqno(3.5)
$$
we obtain $\chi= +\frac{3}{2}$ and 
$\chi= -\frac{3}{2}$, respectively.  
Since we consider that the value of $\chi$ 
must be $\pm \frac{1}{2}$, we consider 
that those states (3.5) are unphysical. 
Therefore, we can remove such 
unwelcome states (exotic states) (3.5). 

This is not only to remove unwelcome states,
but also our model makes anomaly free 
for the SU(5) flavor. 
(See Table 3.)
However,  this choice does not 
make anomaly free for SU(3) family. 
We consider that SU(3) family is not gauge 
symmetry, but "apparent" symmetry. 
In fact, we know the family symmetry is 
badly broken even if it is "symmetry". 

 \begin{table}
\begin{center}
\begin{tabular}{|c|c|cc|} \hline
SU(5)$\times$SU(3) states & $\chi$ & 
$A_5$ & $A_3$ \\ \hline 
 ({\bf 5}{\bf 5}{\bf 5})  =($\overline{\bf 10}$, {\bf 1}) 
&  $+\frac{3}{2}$ & $-1 \times 1$ & 0 \\
 ({\bf 5}{\bf 5}{\bf 3}) =({\bf 10}, {\bf 3})
& $+\frac{1}{2}$ & $+1 \times 3$ &  $+1 \times 10$ \\
({\bf 5}{\bf 3}{\bf 3}) =({\bf 5}, $\overline{\bf 3}$)
& $-\frac{1}{2}$ & $+1 \times 3$ & $ -1 \times 5$ \\ 
 ({\bf 3}{\bf 3}{\bf 3}) =({\bf 1}, {\bf 1})
 & $-\frac{3}{2}$ & 0 & 0 \\
\hline
\end{tabular}

 \caption{\bf Anomaly table for SU(5)$\times$SU(3)} 
\end{center}
\end{table}  
  

\vspace{5mm}
{\large\bf 4. Concluding remarks}

In conclusion, if we take that number of families 
is three, we can easily accept the picture 
that the quarks and leptons are members of 
$(\overline{{\bf 10}} + {\bf 5}, {\bf 3})_L$ of 
SU(5)$_{GUT}\times$SU(3)$_{family}$, i.e. 
$({\bf 10}, {\bf 3})_R$ + 
$( {\bf 5}, \overline{{\bf 3}})_L$. 


In this paper, we have proposed a composite model 
for quarks and leptons where those are given by 
three body states of preons with {\bf 5} of 
SU(5)-flavor and {\bf 3} of SU(3)-family. 
In this model, we would like to emphasis that 
family number is three as seen in Eq.(2.6). 
Of course, in this composite model, it is important
that those preons are confined by the SU(3) hyper-color.

For flavor symmetry, we have taken a picture of 
SU(5)-GUT.  
Regrettably, at present, we do not have any evidence 
for the proton decay. 
However, considering our success of the unified description 
of our quarks and leptons based on the composite 
model,  we have recognized the SU(5) GUT picture 
and we consider that SU(5)-GUT phenomena will be  
confirmed in future experiments.

 The author previously have discussed a similar 
 model  under the 
SU(5)$_{GUT} \times$SU(3)$_{family}$ \cite{YK}.  
However, because the author took a conventional 
field theoretical approach, we obtained unwelcome 
states $(\overline{{\bf 10}},{\bf 1})$ and 
$({\bf 1}, {\bf 1})$ which correspond to 
the states ({\bf 5} {\bf 5} {\bf 5}) and 
({\bf 3} {\bf 3} {\bf 3}) in Table 3 
in addition to the states
given in Table 2. 
In this paper, we adopt a composite picture, 
and we put a selection rule on the chirality
in the composite system (Sec.3). 
This rule is somewhat forcible.
However, with the help of this selection rule,  
we can obtain the desired quarks and lepton 
states only.
In future, the selection rule given in Sec.3 
will be replaced by more reasonable theory. 

It is likely picture that quarks and 
leptons are composite states which are 
consisted of further fundamental constituents
"preons".

%

\vspace{5mm} 

{\large\bf Acknowledgments} 

\vspace{2mm}

The author is grateful to  Prof. T. Yamashita 
for his helpful comments.
This work is supported by JPS KAKENHI Grant 
number JP1903826.




\vspace{5mm}


\begin{thebibliography}{99} 
%
%
\bibitem{quark} M. Gell-Mann, Phys. Lett. {\bf 8}, 214 (1964), 
and G Zeig, CERN preprint 8419/TH412 (1964). 
%
\bibitem{SU5} H. Gergiai and S. L. Grashow, 
Phys. Rev. Lett. {\bf 32}, 438 (1974).
%
\bibitem{YK}  Y. Koide, Phys. Lett. {\bf B797} 134909 (2019).
%
\bibitem{preon} J. C. Pati and A. Salam, Phys. Rev. 
{\bf D 10 }  275 (1974).  

  


\end{thebibliography}
\end{document}